\documentclass[a4paper]{article}
\usepackage[utf8]{inputenc}

\usepackage{multicol}
\usepackage{amsthm}
\usepackage{ulem}
\usepackage{bm}
\usepackage{physics}

\usepackage[dvipdfmx]{graphicx}

\usepackage{bmpsize}
\usepackage{amssymb}
\usepackage{amsmath}
\usepackage{color}
\usepackage{amsthm}
\usepackage{comment}
\usepackage{graphics}
\usepackage{caption}
\usepackage{subcaption}

\newcommand{\BA}{\begin{eqnarray}}
\newcommand{\EA}{\end{eqnarray}}

\definecolor{dgreen}{rgb}{0.0, 0.5, 0.0}

\begin{document}

\fontsize{14pt}{16.5pt}\selectfont

\begin{center}
\bf{Piecewise linear cusp bifurcations \\
in ultradiscrete dynamical systems}
\end{center}

\fontsize{12pt}{11pt}\selectfont
\begin{center}
Shousuke Ohmori$^{1,*)}$ and Yoshihiro Yamazaki$^{2)}$\\ 
\end{center}

\ \\
\textit{
1)~Department of Economics, Hosei University, Machida-shi, Tokyo 194-0298, Japan.\\
2)~Department of Physics, Waseda University, 
    Shinjuku, Tokyo 169-8555, Japan.
}

\bigskip

\noindent
*corresponding author: 42261timemachine@ruri.waseda.jp\\
~~\\
\rm
\fontsize{11pt}{14pt}\selectfont\noindent

\baselineskip 30pt

\noindent
{\bf Abstract}\\
We investigate the dynamical properties of cusp bifurcations in max-plus dynamical systems derived from continuous differential equations through the tropical discretization and the ultradiscrete limit.
A general relationship between cusp bifurcations in continuous and corresponding discrete systems is formulated as a proposition.
For applications of this proposition, we analyze the Ludwig and Lewis models, elucidating the dynamical structure of their ultradiscrete cusp bifurcations obtained from the original continuous models. 
In the resulting ultradiscrete max-plus systems, the cusp bifurcation is characterized by piecewise linear representations, and its behavior is examined through the graph analysis.
\ \\
%



\section{Introduction}
\label{sec:1}

Nonlinear and nonequilibrium phenomena have long been described using two major mathematical frameworks: continuous differential models and discrete difference models\cite{Strogatz, Boccara}. 
Given their differences in formulation, the relationship between continuous and discrete models has remained an intriguing and active area of research\cite{Wolfram}, particularly in exploring how discrete models can retain the essential dynamical features of their continuous counterparts.
A particularly successful approach in this context is ultradiscretization, which serves as a systematic procedure for deriving max-plus equations from continuous or difference equations. 
One notable result is the ultradiscretization of the Korteweg-de Vries equation, leading to the ultradiscrete Lotka-Volterra eq. and the box-ball cellular automaton system, 
which retain the dynamical structures of solitons in the original continuous system\cite{Tokihiro}. 
This connection among continuous, discrete, and ultradiscrete models 
not only provides a novel perspective on integrable systems 
but also shows that the essential dynamical property, 
solitary wave propagation, is retained under ultradiscretization.
Beyond the integrable systems, ultradiscretization has been applied to a wide variety of non-integrable and dissipative systems, including reaction-diffusion systems\cite{Murata, Matsuya, Ohmori2016} and biological models such as inflammatory response networks\cite{Carstea, Willox}. 
These applications highlight the versatility and universality of the ultradiscretization framework in capturing key dynamical behaviors even in non-integrable and dissipative systems.

Recently, we have studied ultradiscretization
of bifurcation phenomena represented 
by nonlinear dynamical systems in one dimension\cite{Ohmori2020,Ohmori2023} and in two dimension\cite{Yamazaki2021,Ohmori2021,Ohmori2022,Ohmori2023_2,Yamazaki2023,Ohmori2023_NF,Ohmori2024,Yamazaki2024}.
%
%
%
%
%
%
%
%
Here we focus on the following 
one-dimensional differential equation of $x=x(t)$,  
	\begin{eqnarray}
    \displaystyle\frac{dx}{dt} = F(x,\alpha)=f(x,\alpha)-g(x,\alpha),
		\label{eqn:ODE}
	\end{eqnarray}
where $f$ and $g$ are positive smooth functions and $\alpha (>0)$ represents the bifurcation parameter.
Employing the tropical discretization\cite{Murata} 
for eq.(\ref{eqn:ODE}), we obtain 
%
%
%
\begin{eqnarray}
    x_{n+1} = F_\tau(x_n,\alpha)
    = \frac{x_n+\tau f(x_n,\alpha)}{x_n+\tau g(x_n,\alpha)} x_n, 
    \label{eqn:TDE}
\end{eqnarray}
where $n$ is the iteration step 
and $\tau$ shows the time interval.
Note that eq.(\ref{eqn:TDE}) is identical 
to eq.(\ref{eqn:ODE}) in the limit of $\tau \to 0$.

The general relationship between the dynamical properties 
of the fixed points obtained respectively 
from eq.(\ref{eqn:ODE}) and eq.(\ref{eqn:TDE}) 
has been reported in our previous paper\cite{Ohmori2023}. 
We have shown that when eq.(\ref{eqn:ODE}) possesses 
saddle node, transcritical, and supercritical pitchfork bifurcations 
at the bifurcation point $(\bar{x},\bar{\alpha})$, 
eq.(\ref{eqn:TDE}) can exhibit them at the same bifurcation point.
Furthermore, we have found that 
these bifurcations can be also retained 
in the ultradiscretized max-plus forms.
%
The mathematical conditions necessary for the preservation of these bifurcations through tropical discretization and ultradiscretization have been discussed in our previous works\cite{Ohmori2023, Ohmori2023_2, Yamazaki2023}.
%
%
%

Until now, studies on the bifurcations of ultradiscrete equations 
and their correspondence with differential and difference equations 
have been conducted only for one-parameter systems.
In this paper, we extend our previous treatment  
to include the cusp bifurcation, 
which is a typical example of a two-parameter bifurcation system.
\section{Cusp bifurcation condition}
\label{sec:2}

For the cusp bifurcation, we consider the two-parameter case 
in eqs.(\ref{eqn:ODE}) and (\ref{eqn:TDE}):
\begin{eqnarray}
    \alpha=(\alpha_1,\alpha_2),
    \;\;\text{where }\;\; \alpha_1,\alpha_2>0.
    \label{eqn:parameter}
\end{eqnarray}
At the bifurcation point $(x,\alpha)=(\bar x, \bar \alpha)$, 
where $\bar\alpha =(\bar\alpha_1,\bar\alpha_2)$, 
$F(\bar x,\bar{\alpha})=0$ is satisfied, 
and the following relations are obtained 
regarding $F(\bar{x},\bar{\alpha})$ and $F_\tau (\bar{x}, \bar{\alpha})$:  
\begin{eqnarray}
    \displaystyle \frac{\partial F_\tau (\bar x, \bar{\alpha})}{\partial x}
    & = & 1+Z_\tau (\bar x, \bar \alpha)D(\bar x,\bar \alpha), 
    \label{eqn:dftau_dx} \\
    \displaystyle \frac{\partial F_{\tau}(\bar x, \bar \alpha)}{\partial \alpha_i} 
    & = & Z_\tau (\bar x, \bar \alpha)\frac{\partial F(\bar x, \bar \alpha)}{\partial \alpha_i}, \;\;\; (i=1,2),  
    \label{eqn:diff.relations1}
\end{eqnarray}
where 
\begin{eqnarray}
    Z_\tau (\bar x, \bar \alpha)
    & = & \frac{\tau \bar x}{\bar x +\tau f(\bar x, \bar \alpha)},
    \label{eqn:Z-function} \\
    D(\bar{x},\bar \alpha)
    & = & \frac{\partial f}{\partial x}(\bar{x},\bar \alpha)-\frac{\partial g}{\partial x}(\bar{x},\bar \alpha).
    \label{eqn:D-function}
\end{eqnarray}
In eq.(\ref{eqn:Z-function}), 
$Z_\tau(\bar x,\bar \alpha)>0$ always holds, 
since $\bar{x}$, $f$, and $\tau$ are positive. 
In the case where eq.(\ref{eqn:ODE}) has the nonhyperbolic fixed point, 
$F(\bar x,\bar{\alpha})=0$ 
and $\displaystyle\frac{\partial F(\bar x, \bar \alpha)}{\partial x}=D(\bar{x},\bar{\alpha})=0$ hold, 
and we obtain 
\begin{eqnarray}
    \displaystyle\frac{\partial^2 F_{\tau}(\bar x, \bar \alpha)}{\partial^2 x} = Z_\tau (\bar x, \bar \alpha)\frac{\partial^2 F(\bar x, \bar \alpha)}{\partial^2 x}
    \label{eqn:diff.relations2}
\end{eqnarray}
and
\begin{eqnarray}
    & & \displaystyle\frac{\partial F_{\tau}(\bar x, \bar \alpha)}{\partial \alpha_1}\displaystyle\frac{\partial^2 F_{\tau}(\bar x, \bar \alpha)}{\partial x\partial \alpha_2}-\displaystyle\frac{\partial F_{\tau}(\bar x, \bar \alpha)}{\partial \alpha _2}\displaystyle\frac{\partial^2 F_{\tau}(\bar x, \bar \alpha)}{\partial x\partial \alpha_1}\nonumber \\
    &  & =
    Z_\tau ^2(\bar x, \bar \alpha)\left(   
    \displaystyle\frac{\partial F(\bar x, \bar \alpha)}{\partial \alpha_1}\displaystyle\frac{\partial^2 F(\bar x, \bar \alpha)}{\partial x\partial \alpha_2}-\displaystyle\frac{\partial F(\bar x, \bar \alpha)}{\partial \alpha _2}\displaystyle\frac{\partial^2 F(\bar x, \bar \alpha)}{\partial x\partial \alpha_1}\right).
    \label{eqn:diff.relations4}
\end{eqnarray}
Furthermore, when $\displaystyle\frac{\partial^2 F(\bar x, \bar \alpha)}{\partial^2 x}=0$ is satisfied, we obtain  
\begin{eqnarray}
    \displaystyle\frac{\partial^3 F_\tau(\bar x, \bar \alpha)}{\partial x^3} = Z_\tau (\bar x, \bar \alpha)\frac{\partial^3 F(\bar x, \bar \alpha)}{\partial x^3}.
    \label{eqn:diff.relation3}
\end{eqnarray}
Based on eqs.(\ref{eqn:diff.relations2})-(\ref{eqn:diff.relation3}),  
the following proposition about the cusp bifurcation holds.
\begin{description}
	\item[Proposition: cusp bifurcation condition]
		\ \\
		When eq.(\ref{eqn:ODE}) satisfies the condition 
		for the continuous cusp bifurcation 
        at the bifurcation point $(\bar x, \bar \alpha)$, which is 
        \[ 
          \displaystyle F(\bar x, \bar \alpha) = 0, 
		 \displaystyle \frac{\partial F(\bar x, \bar \alpha)}{\partial x}=0, 
          \displaystyle \frac{\partial^2 F(\bar x, \bar \alpha)}{\partial x^2} =0, 
          \displaystyle \frac{\partial^3 F(\bar x, \bar \alpha)}{\partial x^3} \not=0, \text{ and}
        \]
        \[
          \displaystyle \frac{\partial F(\bar x, \bar \alpha)}{\partial \alpha_1}\frac{\partial^2 F(\bar x, \bar \alpha)}{\partial x\partial \alpha_2}-\frac{\partial F(\bar x, \bar \alpha)}{\partial \alpha _2}\frac{\partial^2 F(\bar x, \bar \alpha)}{\partial x\partial \alpha_1} \not =0, 
        \]
		eq.(\ref{eqn:TDE}) also satisfies 
        the discrete cusp bifurcation condition \cite{Kuznetsov2023}
        at the cusp bifurcation point $(\bar x,\bar \alpha)$, 
        which is given as 
        \[ 
          \displaystyle F_\tau(\bar x, \bar \alpha) = \bar x, 	  \displaystyle \frac{\partial F_\tau(\bar x, \bar \alpha)}{\partial x}=1,  
          \displaystyle \frac{\partial^2 F_\tau(\bar x, \bar \alpha)}{\partial x^2} =0,  
          \displaystyle \frac{\partial^3 F_\tau(\bar x, \bar \alpha)}{\partial x^3} \not=0, \text{ and}
        \]
        \[ 
          \displaystyle \frac{\partial F_{\tau}(\bar x, \bar \alpha)}{\partial \alpha_1}\frac{\partial^2 F_{\tau}(\bar x, \bar \alpha)}{\partial x\partial \alpha_2}-\frac{\partial F_{\tau}(\bar x, \bar \alpha)}{\partial \alpha _2}\frac{\partial^2 F_{\tau}(\bar x, \bar \alpha)}{\partial x\partial \alpha_1} \not =0.
        \]
\end{description}
In the case of $\tau \to \infty$, 
eqs.(\ref{eqn:TDE}) and (\ref{eqn:Z-function}) become  
\begin{eqnarray}
    x_{n+1}=x_n\frac{f(x_n,\alpha)}{g(x_n,\alpha)}
    \label{eqn:TDE_tau_inf}
\end{eqnarray}
and 
\begin{eqnarray}
    Z_\infty (\bar x, \bar \alpha)=\frac{\bar x}{f(\bar x, \bar \alpha)}, 
    \label{eqn:Z-function_infinity}
\end{eqnarray}
respectively.
Note that the above proposition 
for the cusp bifurcation condition holds 
even in the case of $\tau \to \infty$.
In other words, the cusp bifurcation 
of the original continuous dynamical system is retained  
in the discrete dynamical system 
obtained by the tropical discretization for any $\tau$.

\section{Application}
\label{sec:3}

\begin{figure}[h!]
  \begin{minipage}[b]{0.5\linewidth}
    \centering
    \includegraphics[width=0.95\linewidth]{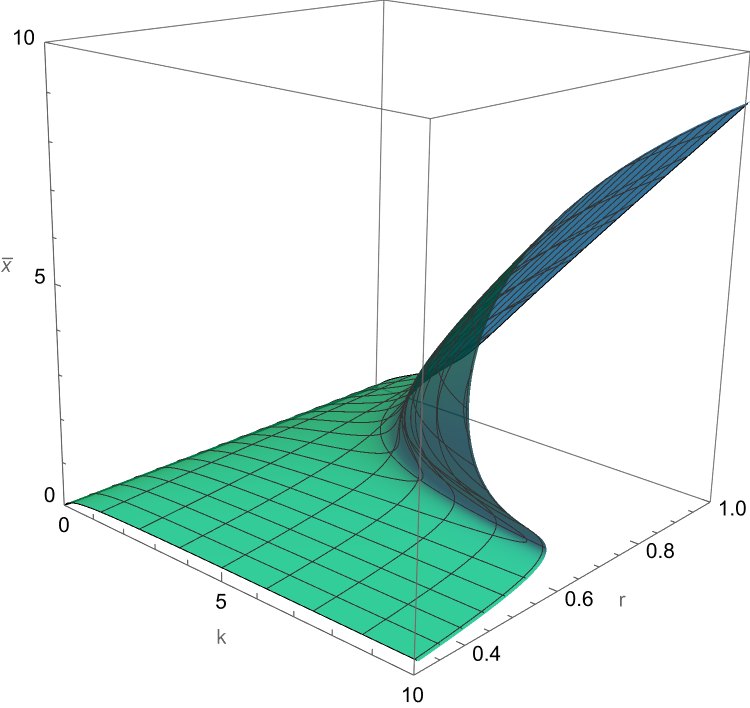}
    \subcaption{ }
  \end{minipage}
  \begin{minipage}[b]{0.5\linewidth}
    \centering
    \includegraphics[width=0.95\linewidth]{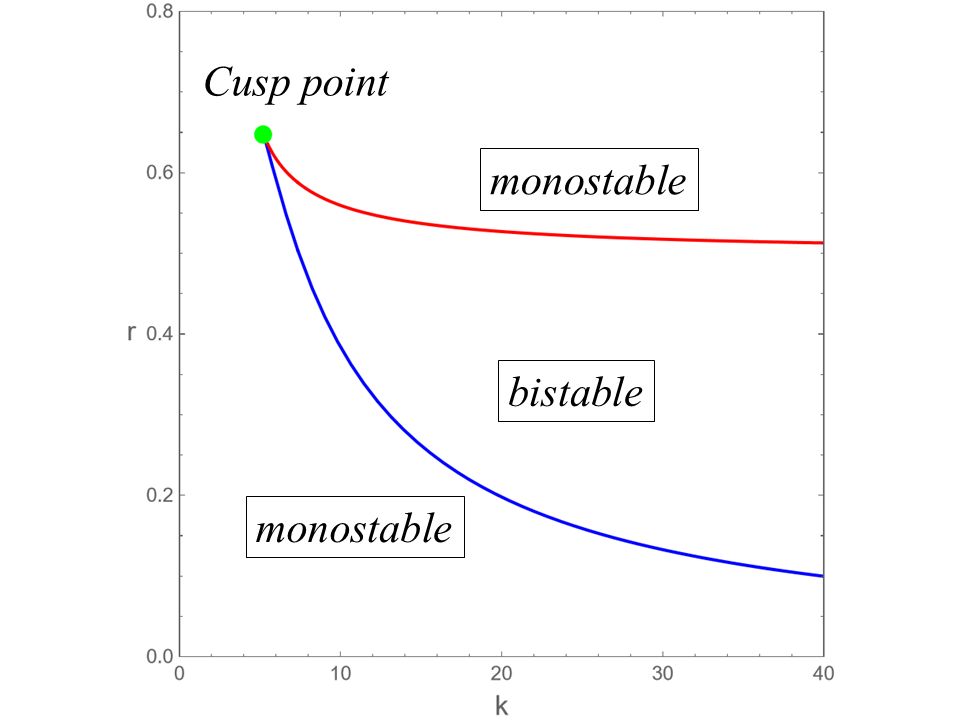}
    \subcaption{ }
  \end{minipage}
  \caption{ 
    (a) The cusp catastrophe surface of eq.(\ref{eqn:fp_intersection}).
    (b) The bifurcation curve of eq.(\ref{eqn:bif_curve}).
    The cusp point is $\bar{x} = \sqrt{3}$ and  $(\bar{r}, \bar{k})=\left(\displaystyle\frac{3\sqrt{3}}{8},3\sqrt{3}\right)$.
  }
  \label{Fig.Conti_catas}
\end{figure}

For application of the above proposition, 
we now focus on the Ludwig model, 
which is known as an ecological model with the cusp bifurcation\cite{Strogatz,Ludwig}. 
The Ludwig model is given by the following one-dimensional continuous dynamical system: 
	\begin{eqnarray}
    \displaystyle\frac{dx}{dt}
    =F(x, r, k) = rx\left(1-\frac{x}{k}\right )-\frac{x^2}{1+x^2}, 
		\label{eqn:Ludwig}
	\end{eqnarray}
where $x > 0$ and $r,k>0$.
The fixed point of eq.(\ref{eqn:Ludwig}), $\bar{x}$, 
satisfies $F(\bar{x}, r, k)=0$, i.e., 
\begin{eqnarray}
    \displaystyle r \left(1-\displaystyle\frac{\bar{x}}{k}\right)=\displaystyle\frac{\bar{x}}{1+\bar{x}^2}. 
    \label{eqn:fp_intersection}
\end{eqnarray}
Furthermore by considering $\displaystyle \frac{\partial F(\bar x, r, k)}{\partial x}=0$, 
we obtain the following bifurcation curve for eq.(\ref{eqn:Ludwig}): 
\begin{eqnarray}
    r  =  \displaystyle\frac{2 \bar{x}^3}{\left(1+\bar{x}^2\right)^2}, 
    \;\;\;\;\;
    k  =  \frac{2\bar{x}^3}{\bar{x}^2-1}.
    \label{eqn:bif_curve}
\end{eqnarray}
Since the cusp bifurcation point ($\bar{x}$, $\bar{r}$, $\bar{k}$) 
also satisfies 
$\displaystyle \frac{\partial^{2} F(\bar x, \bar r, \bar k)}{\partial x^{2}}=0$, 
we obtain 
\begin{equation}
    \bar{x}=\sqrt{3},  \;\;\; 
    (\bar{r},\bar{k})=\left(\displaystyle\frac{3\sqrt{3}}{8},3\sqrt{3}\right).
    \label{eqn:bp_Ludwig}
\end{equation}
At this bifurcation point, it is also confirmed that 
$\displaystyle  \frac{\partial^3 F(\bar x, \bar r, \bar k)}{\partial x^3} = -\frac{3\sqrt{3}}{16} \ne 0$, 
and $\displaystyle \frac{\partial F(\bar x, \bar r, \bar k)}{\partial r}\frac{\partial^2 F(\bar x, \bar r, \bar k)}{\partial x\partial k}-\frac{\partial F(\bar x, \bar r, \bar k)}{\partial k}\frac{\partial^2 F(\bar x, \bar r, \bar k)}{\partial x\partial r} = \frac{1}{8\sqrt{3}} \ne 0$.
Figure \ref{Fig.Conti_catas} shows (a) the cusp catastrophe surface of eq.(\ref{eqn:fp_intersection}) 
and (b) the cusp bifurcation curve of eq.(\ref{eqn:bif_curve}).
Based on eq.(\ref{eqn:TDE}), the tropical discretization 
of eq.(\ref{eqn:Ludwig}) brings about 
\begin{eqnarray}
    x_{n+1}=\displaystyle \frac{x_n+\tau r x_n}{1+\tau\left( \displaystyle\frac{r}{k}x_n+\frac{x_n}{1+x_n^2}\right)},
    \label{eqn:tropical_Ludwing}
\end{eqnarray}
where we set  
\begin{eqnarray}
    f(x,r,k)=rx~~
    \text{and}~~
    g(x,r,k)=\displaystyle\frac{rx^2}{k}+\frac{x^2}{1+x^2}.
    \label{eqn:f_and_g}
\end{eqnarray}
From the above proposition, eq.(\ref{eqn:tropical_Ludwing}) 
also possesses the cusp bifurcation.
It is confirmed that the fixed point of eq.(\ref{eqn:ODE}) 
is identical to that of eq.(\ref{eqn:TDE})\cite{Ohmori2023}.
Then the fixed point of eq.(\ref{eqn:tropical_Ludwing}), 
which is also $\bar{x}$, satisfies eq.(\ref{eqn:fp_intersection}).
%
%
%
In addition, according to ref.\cite{Ohmori2023}, 
the stability of a fixed point $\bar x$ for eq.(\ref{eqn:TDE}) depends on the sign of $\kappa(\bar{x})$, which is defined as 
\begin{eqnarray}
    \kappa(\Bar{x})  =  -\displaystyle\frac{2\Bar{x}}{\Bar{x}D(\Bar{x})+2f(\Bar{x})}.
\label{eqn:Kappa}
\end{eqnarray}
We have confirmed that if $\bar{x}$ for eq.(\ref{eqn:ODE}) is unstable, 
then $\bar x$ for eq.(\ref{eqn:TDE}) is also unstable.
And if $\bar x$ for eq.(\ref{eqn:ODE}) is 
stable and $\kappa(\Bar{x})<0$, 
then $\bar x$ for eq.(\ref{eqn:TDE}) 
is also stable for any $\tau >0$.
%
%
In the case of eq.(\ref{eqn:tropical_Ludwing}), 
$\kappa(\bar{x})$ can be calculated as  
\begin{eqnarray}
    \kappa(\Bar{x}) & = & -\displaystyle\frac{2\Bar{x}}{\Bar{x}D(\Bar{x})+2f(\Bar{x},r,k)}\nonumber \\
    & = & -\frac{2}{D(\bar{x})+2r} 
    = -\frac{2}{r+\displaystyle \frac{2\bar{x}^3}{(1+\bar{x}^{2})^{2}}} < 0.
\label{eqn:D_xbar}
\end{eqnarray}
%
%
%
%
%
%
%
%
Therefore, the stability of $\bar x$ 
in eq.(\ref{eqn:Ludwig}) is retained 
in eq.(\ref{eqn:tropical_Ludwing}) for any $(r,k)$ and $\tau$.
%
%
%
Furthermore, from the above proposition, eq.(\ref{eqn:tropical_Ludwing}) exhibits the cusp bifurcation at the bifurcation point $(\bar{x},\bar{r},\bar{k})$ given in eq.(\ref{eqn:bp_Ludwig}).
The bifurcation curves for eq.(\ref{eqn:tropical_Ludwing}) 
are also given by eq.(\ref{eqn:bif_curve}). 
Note that eq.(\ref{eqn:bif_curve}) obtained from eq.(\ref{eqn:tropical_Ludwing}) is independent of $\tau$.
%
%
%
%
%
%
%
%
%
%
%
Here we note that eq.(\ref{eqn:tropical_Ludwing}) becomes
\begin{eqnarray}
    x_{n+1}=\displaystyle \frac{r}{\displaystyle\frac{r}{k}+\frac{1}{1+x_n^2}}
    \label{eqn:tropical_Ludwing_limit}
\end{eqnarray}
in the limit of $\tau \to \infty$. 
Even in this limit for $\tau$, 
eq.(\ref{eqn:tropical_Ludwing_limit}) has 
the cusp bifurcation and the bifurcation curves are given 
by eq.(\ref{eqn:bif_curve}).
%
%
%
\section{Ultradiscretization}

\begin{figure}[h!]
  \begin{minipage}[b]{0.5\linewidth}
    \centering
    \includegraphics[width=0.95\linewidth]{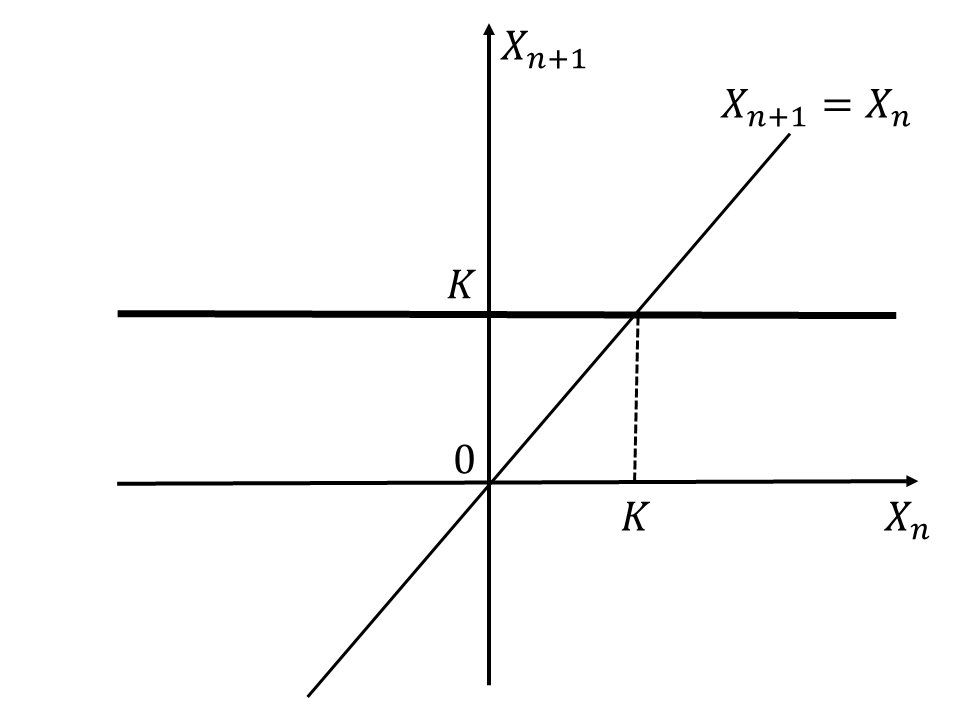}
    \subcaption{ }
  \end{minipage}
  \begin{minipage}[b]{0.5\linewidth}
    \centering
    \includegraphics[width=0.95\linewidth]{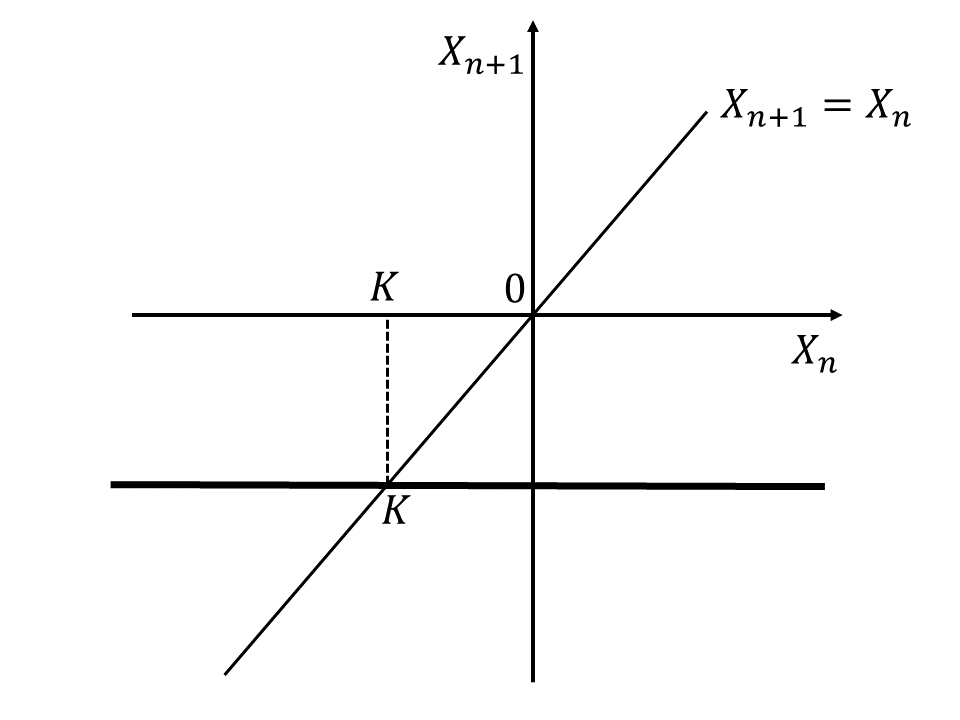}
    \subcaption{ }
  \end{minipage}
  \caption{ The graphs of eq.(\ref{eqn:UD_Ludwing}) for $R>K$. 
    (a) $K>0$ and (b) $K<0$.
  }
  \label{Fig.1}
\end{figure}

\begin{figure}[h!]
  \begin{minipage}[b]{0.5\linewidth}
    \centering
    \includegraphics[width=0.95\linewidth]{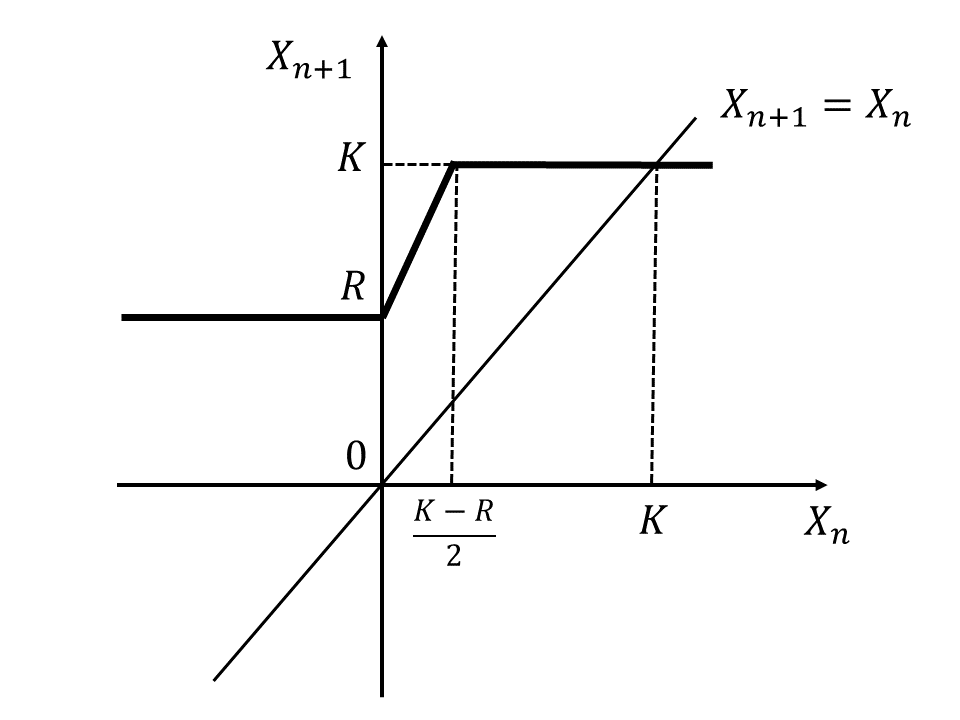}
    \subcaption{ }
  \end{minipage}
  \begin{minipage}[b]{0.5\linewidth}
    \centering
    \includegraphics[width=0.95\linewidth]{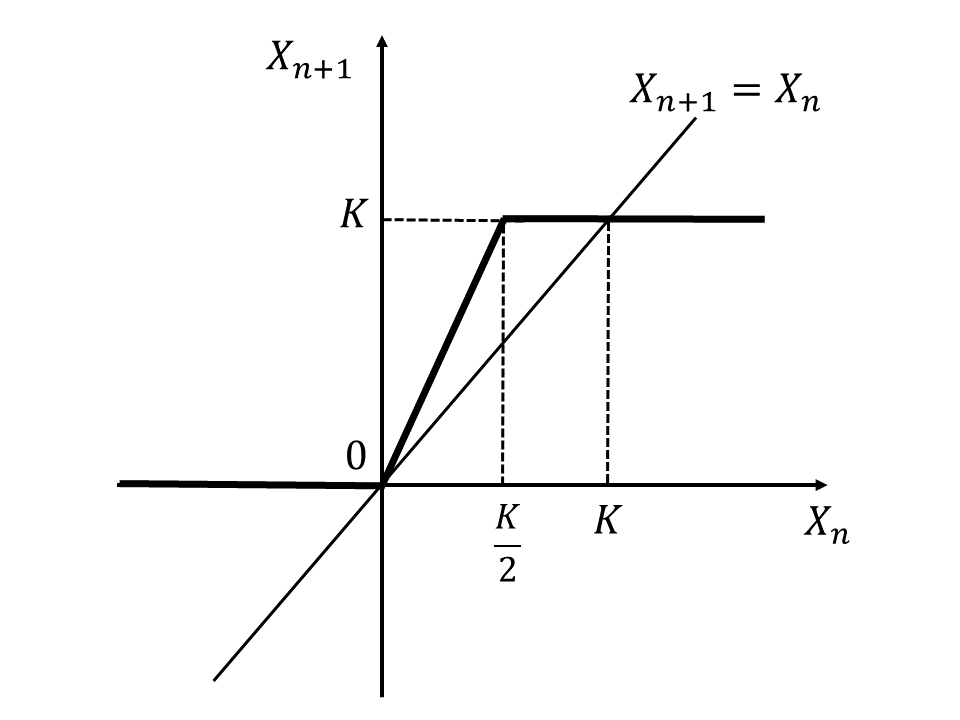}
    \subcaption{ }
  \end{minipage}
  \begin{minipage}[b]{0.5\linewidth}
    \centering
    \includegraphics[width=0.95\linewidth]{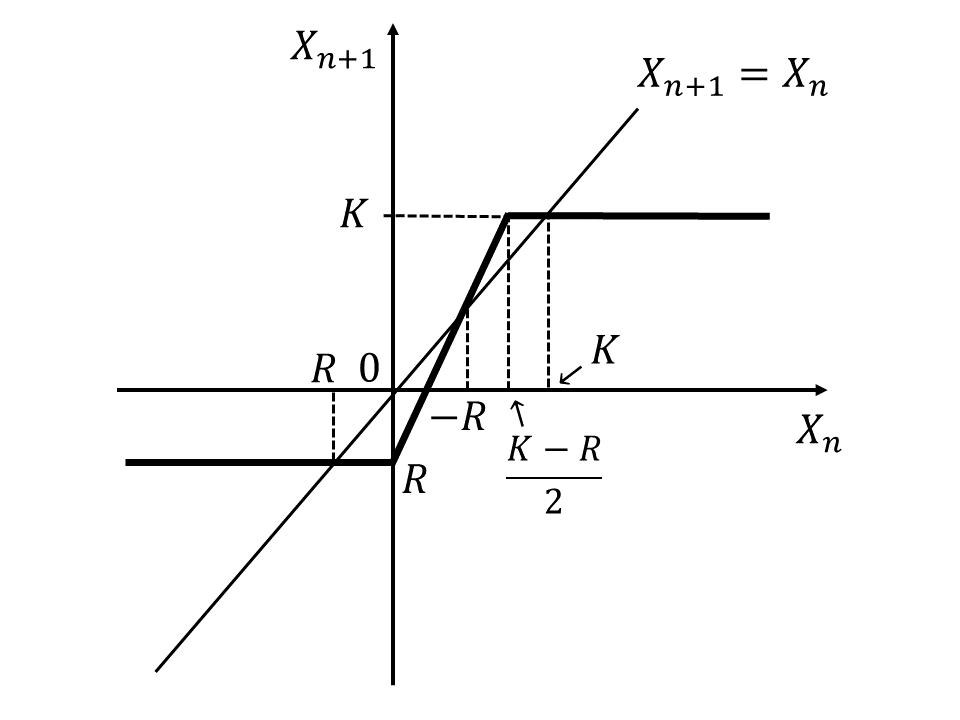}
    \subcaption{ }
  \end{minipage}
  \begin{minipage}[b]{0.5\linewidth}
    \centering
    \includegraphics[width=0.95\linewidth]{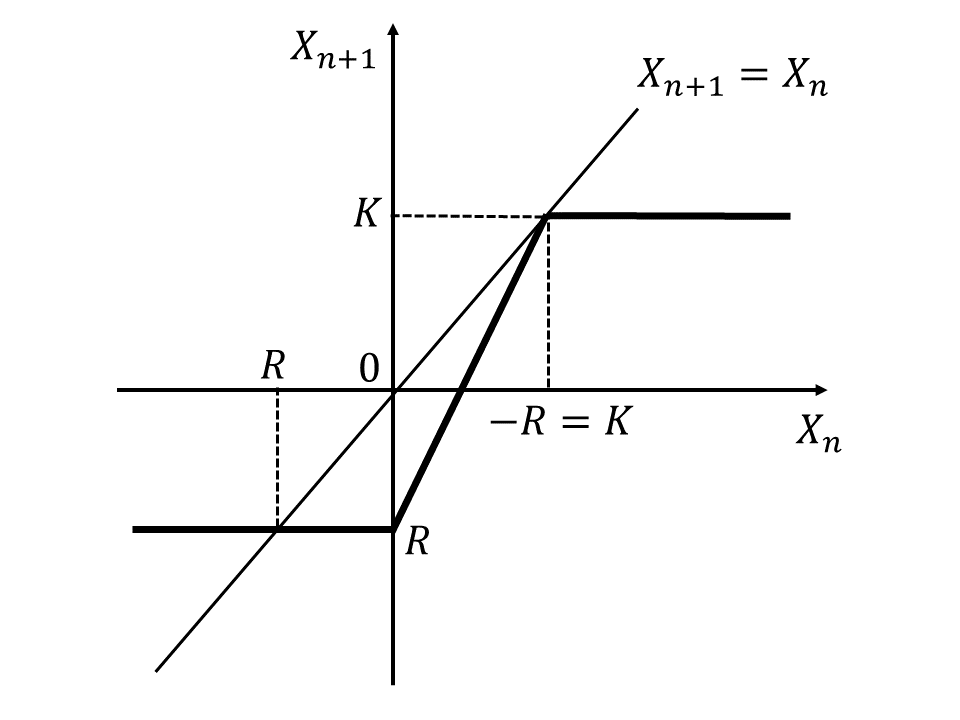}
    \subcaption{ }
  \end{minipage}
  \begin{minipage}[b]{0.5\linewidth}
    \centering
    \includegraphics[width=0.95\linewidth]{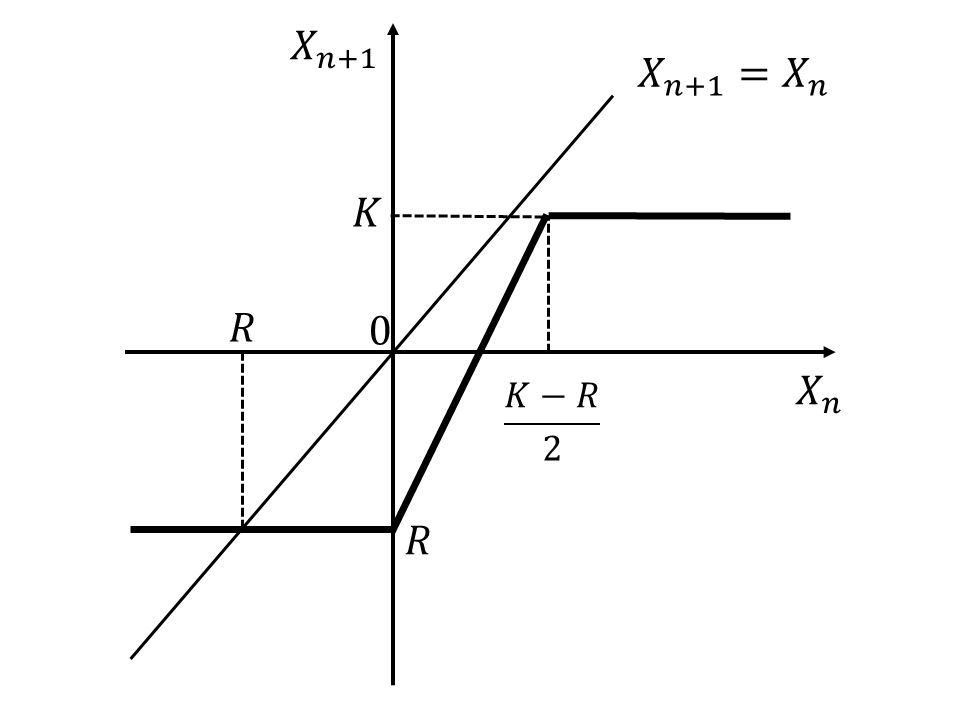}
    \subcaption{ }
  \end{minipage}
  \caption{ The graphs of eq.(\ref{eqn:UD_Ludwing}) 
    for (a) $R>0$, (b) $R=0$, (c) $R<0$ and $K>-R$, 
    (d) $R<0$ and $K = -R$, and (e) $R<0$ and $K<-R$.
  }
  \label{Fig.3}
\end{figure}

\begin{figure}[h!]
  \begin{minipage}[b]{0.5\linewidth}
    \centering
    \includegraphics[width=0.95\linewidth]{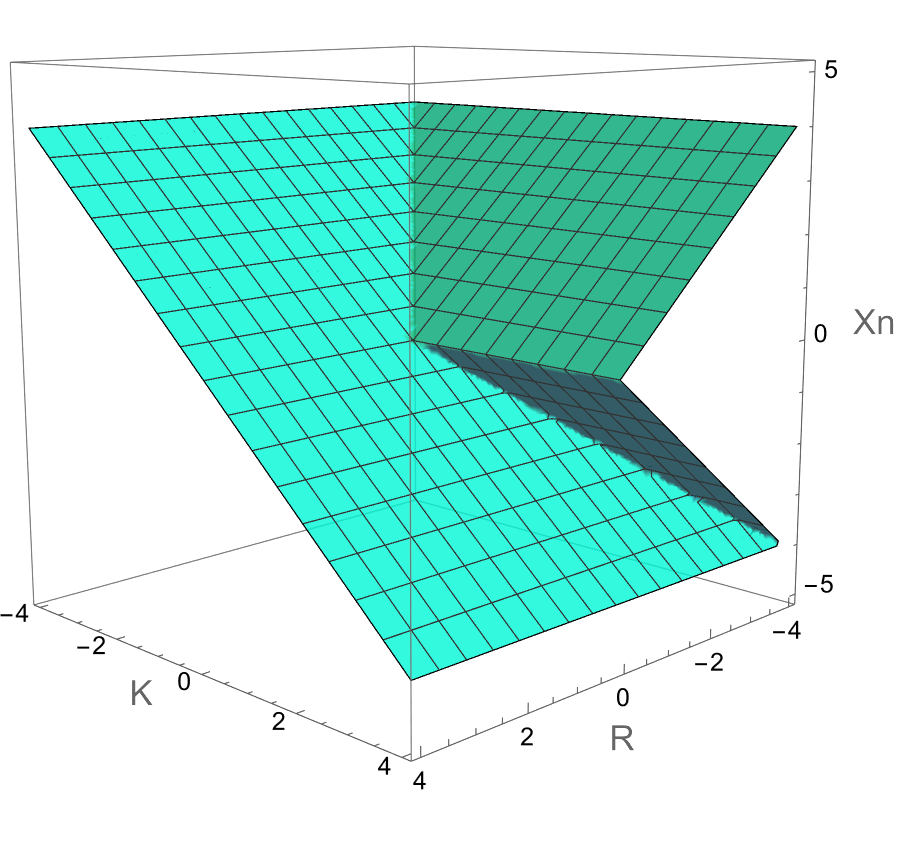}
    \subcaption{ }
  \end{minipage}
  \begin{minipage}[b]{0.5\linewidth}
    \centering
    \includegraphics[width=0.95\linewidth]{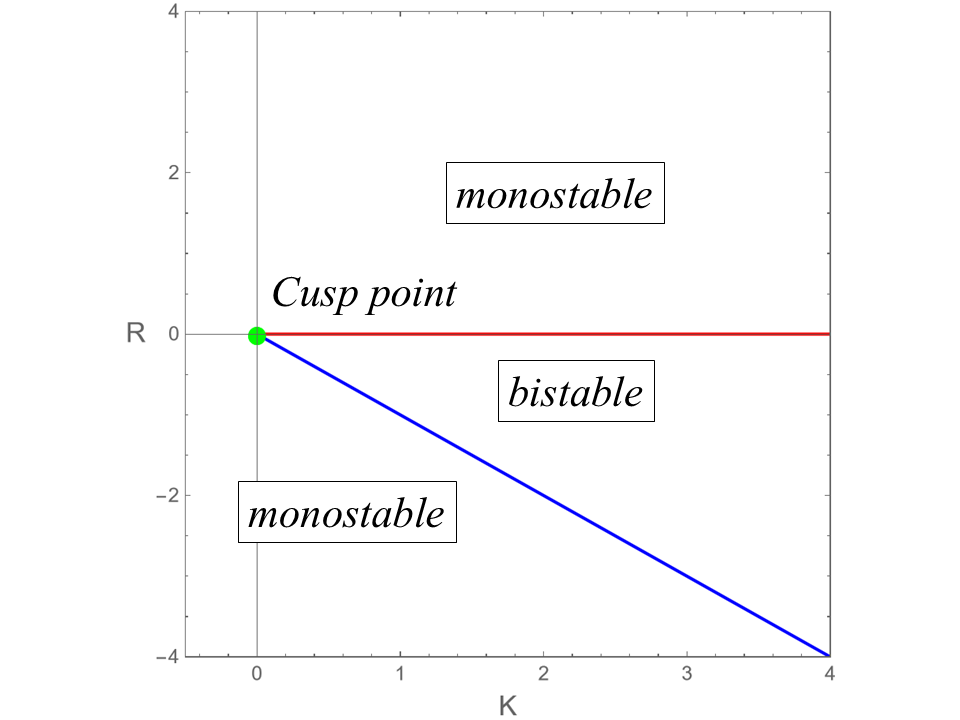}
    \subcaption{ }
  \end{minipage}
  \caption{ (a) The cusp catastrophe surface for eq.(\ref{eqn:UD_Ludwing}). 
    (b) The bifurcation curve for eq.(\ref{eqn:UD_Ludwing}).
    The cusp point is $X_n=0$ at $(R,K)=(0,0)$.
  }
  \label{Fig.4}
\end{figure}

To derive the ultradiscrete max-plus equation for eq.(\ref{eqn:tropical_Ludwing_limit}), 
the variable transformations
\begin{eqnarray}
    x_n=e^{X_n/\varepsilon},~r=e^{R/\varepsilon},~k=e^{K/\varepsilon},
    \label{transformation}
\end{eqnarray}
are applied.
Then, the ultradiscrete limit
\[
\lim_{\varepsilon \to +0}\varepsilon\log(e^{A/\varepsilon}+e^{B/\varepsilon})=\max(A,B)	
\]
provides the following max-plus equation,
%
%
%
%
%
%
\begin{eqnarray}
    X_{n+1}=R-\max(R-K,-\max(0,2X_n)).    
    \label{eqn:UD_Ludwing}
\end{eqnarray}
%
%
The dynamical properties of eq.(\ref{eqn:UD_Ludwing}) 
can be understood by the graph analysis (cobweb plot)\cite{Ohmori2020}.
First we consider the case $R>K$.
Figure \ref{Fig.1} shows the graphs of eq.(\ref{eqn:UD_Ludwing}) for (a) $K>0$ and (b) $K<0$.
Since $\max (0, 2X_{n}) \geq 0$, 
eq.(\ref{eqn:UD_Ludwing}) becomes $X_{n+1} = K$ 
for any $X_{n}$.
Therefore, $X_n=K$ is the unique fixed point 
and is stable.
Next we consider the case $R<K$, where
eq.(\ref{eqn:UD_Ludwing}) can be rewritten as
\begin{equation}
    X_{n+1} = \begin{cases}
        R & \left( X_{n} < 0 \right), \\
        2X_{n} + R & \left( 0 \leq X_{n} < \displaystyle \frac{K-R}{2} \right), \\
        K & \left( X_{n} \geq \displaystyle \frac{K-R}{2} \right).
    \end{cases}
    \label{eqn:UD_Ludwing_2}
\end{equation}
Equation (\ref{eqn:UD_Ludwing_2}) is piecewise linear and can be divided into the five cases according to the type of intersection 
with the identity line $X_{n+1} = X_n$, as shown in Fig.\ref{Fig.3}.
Specifically, Fig.\ref{Fig.3} illustrates the following five cases: 
(a) $R>0$, (b) $R=0$, (c) $R<0$ and $K>-R$, 
(d) $R<0$ and $K = -R$, and (e) $R<0$ and $K<-R$.
(a) For $R > 0$, 
it is clear that $X_n=K$ is a stable fixed point.
(b) For $R=0$, 
eq.(\ref{eqn:UD_Ludwing_2}) intersects the identity line 
at $X_n=K$ and $X_n=0$.
It is found that $X_n=K$ is a stable fixed point 
and $X_n=0$ is a half-stable fixed point.
(c) For $R<0$ and $K>-R$, eq.(\ref{eqn:UD_Ludwing_2}) has 
the three fixed points $X_{n} = K, \pm R$, 
where $X_n = +R,K$ are stable and $X_n = -R$ is unstable.
(d) For $R<0$ and $K = -R$, 
eq.(\ref{eqn:UD_Ludwing_2}) intersects the identity line 
at $X_n=R$ and $X_n=K$.
It is found that $X_n=R$ is a stable fixed point 
and $X_n=K$ is a half-stable fixed point.
(e) For $R<0$ and $K<-R$, 
it is clear that $X_n=R$ is a stable fixed point.
Therefore from above, the dynamical properties 
of eq.(\ref{eqn:UD_Ludwing}) are summarized as follows.
\begin{description}
\item [(1)] When $R>K$, $X_n=K$ is the unique fixed point and is stable.
\item [(2)]	When $R<K$,  
    \begin{description}
        \item[(2-i)] if $R>0$, 
        $X_n=K$ is the unique fixed point that is stable,
        \item[(2-ii)] if $R<0$ and $K>-R$, 
        $X_n=\pm R$ $X_n=K$ are the fixed points.
        $+R$ and $K$ are stable (bistable) and $-R$ is unstable.
        \item[(2-iii)] if $R<0$ and $K<-R$, 
        $X_n=R$ is the unique fixed point that is stable.
    \end{description}

\end{description}
Figure \ref{Fig.4}(a) shows 
the cusp catastrophe surface for eq.(\ref{eqn:UD_Ludwing}), 
which is plotted according to the above summary.
And the cusp bifurcation curve for eq.(\ref{eqn:UD_Ludwing}) is also shown in Fig.\ref{Fig.4}(b).
The cusp point is $X=0$ at $(R,K)=(0,0)$.
A comparison between Figs. \ref{Fig.Conti_catas} and \ref{Fig.4} 
indicates that the features of the cusp bifurcation 
are retained even after piecewise linearization through ultradiscretization.
%
%


\section{Discussion}
\label{sec:5}
\begin{figure}[h!]
  \begin{minipage}[b]{0.5\linewidth}
    \centering
    \includegraphics[width=0.95\linewidth]{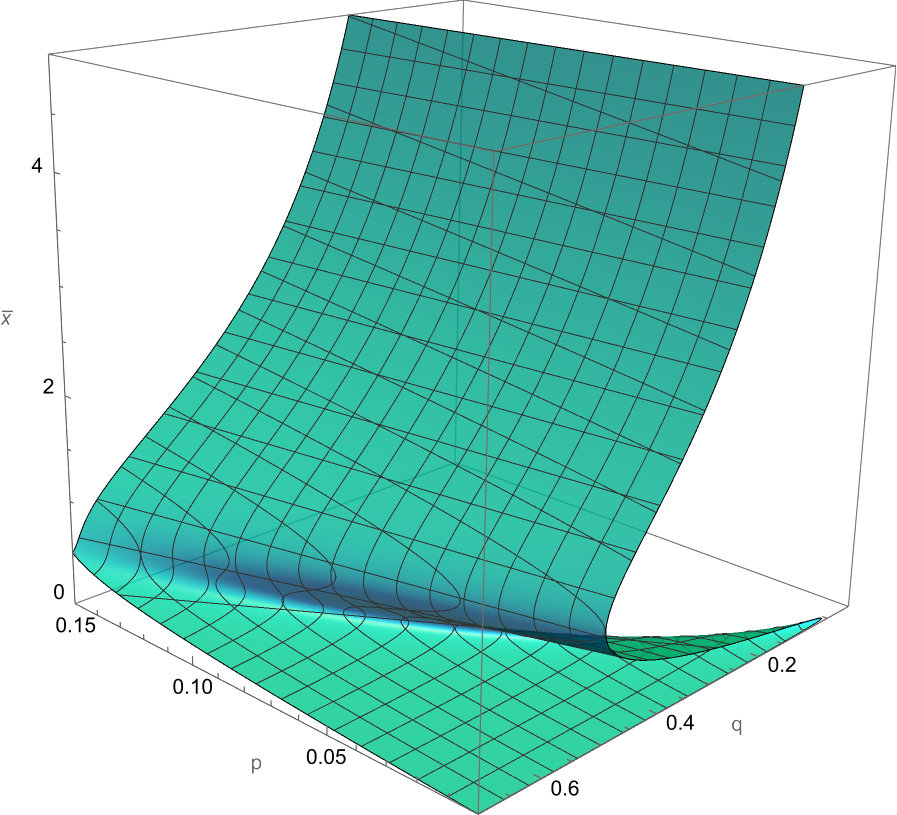}
    \subcaption{ }
  \end{minipage}
  \begin{minipage}[b]{0.5\linewidth}
    \centering
    \includegraphics[width=0.95\linewidth]{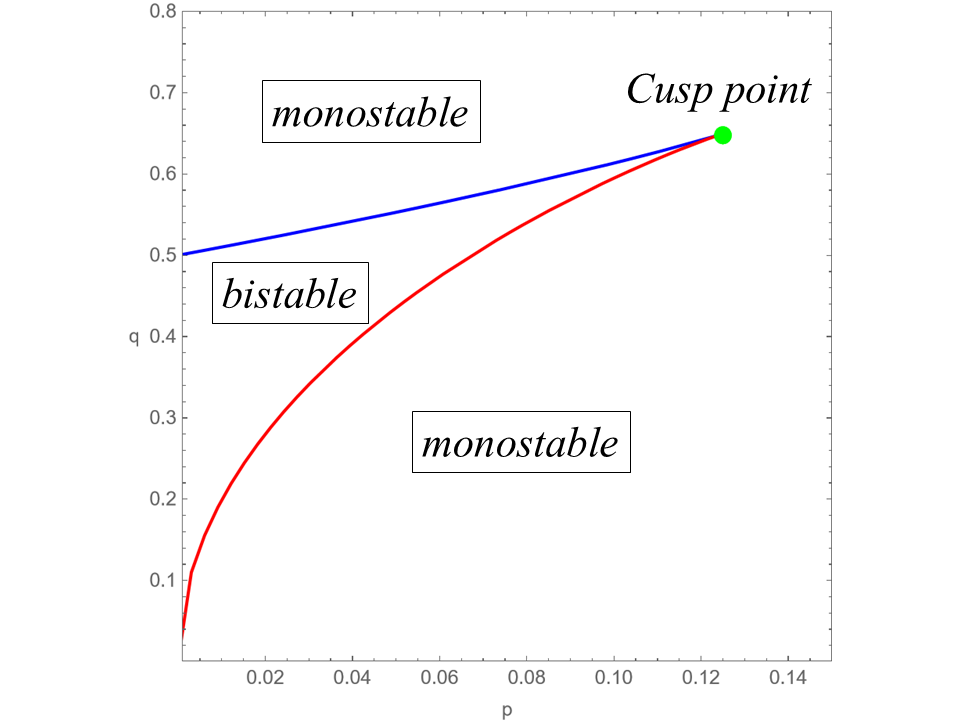}
    \subcaption{ }
  \end{minipage}
  \caption{ (a) The cusp catastrophe surface of eq.(\ref{eqn:fp_intersection_for_Lewis}). 
    (b) The bifurcation curve of eq.(\ref{eqn:bif_curve_for_Lewis}).
    The cusp point is $\displaystyle \bar{x} = \frac{1}{\sqrt{3}}$ at 
    $\displaystyle (\bar{p}, \bar{q})=\left(\frac{1}{8},\frac{3\sqrt{3}}{8}\right)$.
  }
  \label{Fig.Lewis}
\end{figure}

We demonstrate another example.
The following equation is the biochemical model having the cusp bifurcation, proposed by J. Lewis\cite{Strogatz,Lewis}.
\begin{equation}
    \frac{dx}{dt}
    = G(x,p,q) = p-qx+\frac{x^2}{1+x^2}, 
    \label{eqn:Lewis}
\end{equation}
where $x\geq 0$ and $p,q>0$.
Equation (\ref{eqn:Lewis}) possesses the fixed points, $\bar{x}$, 
which satisfy $G(\bar{x}, p, q) = 0$:  
\begin{equation}
    q \bar{x} - p = \frac{\bar{x}^2}{1 + \bar{x}^2}.
    \label{eqn:fp_intersection_for_Lewis}
\end{equation}
Figure \ref{Fig.Lewis}(a) shows the cusp catastrophe surface 
given by eq.(\ref{eqn:fp_intersection_for_Lewis}).
The cusp bifurcation curve are also derived from  
$G(\bar{x}, p, q) = 0$ and 
$\displaystyle \frac{\partial}{\partial x}G(\bar{x}, p, q) = 0$, 
which yield 
\begin{equation}
  p = \frac{(1-\bar{x}^2)\bar{x}^2}{(1+\bar{x}^2)^2},
  \;\;\;\;\; 
  q = \frac{2\bar{x}}{\left(1+\bar{x}^2\right)^2}.
  \label{eqn:bif_curve_for_Lewis}
\end{equation}
The bifurcation curve given by eq. (\ref{eqn:bif_curve_for_Lewis}) 
is shown in Fig. \ref{Fig.Lewis} (b) as functions of $p$ and $q$.
And the cusp point of eq.(\ref{eqn:Lewis}) is 
$\displaystyle \bar{x}=\frac{1}{\sqrt{3}}$ at 
$\displaystyle \left( \bar{p}, \bar{q} \right) = \left( \frac{1}{8},\frac{3\sqrt{3}}{8} \right)$.
Note that eq.(\ref{eqn:Lewis}) satisfies the cusp bifurcation conditions: 
$\displaystyle G(\bar{x},\bar{p},\bar{q})=\frac{\partial G(\bar x, \bar p, \bar q)}{\partial x}=\frac{\partial^2 G(\bar x, \bar p, \bar q)}{\partial x^2}=0$,
$\displaystyle \frac{\partial^3 G(\bar x, \bar p, \bar q)}{\partial x^3} = -\frac{27\sqrt{3}}{16} \ne 0$, 
and $\displaystyle \frac{\partial G(\bar x, \bar p, \bar q)}{\partial p}\frac{\partial^2 G(\bar x, \bar p, \bar q)}{\partial x\partial q}-\frac{\partial G(\bar x, \bar p, \bar q)}{\partial q}\frac{\partial^2 G(\bar x, \bar p, \bar q)}{\partial x\partial p} = -1 \ne 0$.

The discretized equation of eq.(\ref{eqn:Lewis}) can be obtained as follows 
by tropical discretization based on eq.(\ref{eqn:TDE}) 
by setting $f(x, p, q)=p+\displaystyle\frac{x^2}{1+x^2}$ 
and $g(x, p, q)=qx$, 
\begin{eqnarray}
    x_{n+1}=\displaystyle \frac{x_n+\tau\left(p+ \displaystyle\frac{x^2_n}{1+x_n^2}\right) }{1+\tau q }.
    \label{eqn:tropical_Lewis}
\end{eqnarray}
From the above proposition, it is found that 
eq.(\ref{eqn:tropical_Lewis}) also exhibits the cusp bifurcation.
When $\tau \to \infty$, eq.(\ref{eqn:tropical_Lewis}) becomes 
\begin{eqnarray}
    x_{n+1}=\displaystyle\frac{p}{q}+\frac{x^2_n}{q(1+x_n^2)}.
    \label{eqn:tropical_Lewis_limit}
\end{eqnarray}
Note that the fixed points and the bifurcation curve for eq.(\ref{eqn:tropical_Lewis_limit}) coincide 
with those of eq.(\ref{eqn:Lewis}).

Applying the variable transformations,
\begin{eqnarray}
    x_n=e^{X_n/\varepsilon},~p=e^{P/\varepsilon},~q=e^{Q/\varepsilon},
    \label{transformation_Lewis}
\end{eqnarray}
to eq.(\ref{eqn:tropical_Lewis_limit}), 
and taking the ultradiscrete limit,  
the following max-plus equation can be obtained,
\begin{eqnarray}
    X_{n+1} & = & \max(P-Q,2X_n-Q-\max(0, 2X_n)) \notag \\
             & = & -Q+\max(P,-\max(0,-2X_n)).    
    \label{eqn:UD_Lewis_calculation}
\end{eqnarray}
Considering the transformation $X_n \to -X_n$, 
eq.(\ref{eqn:UD_Lewis_calculation}) becomes 
\begin{eqnarray}
    X_{n+1}  =  Q-\max(P,-\max(0,2X_n)).    
    \label{eqn:UD_Lewis}
\end{eqnarray}
Comparing eq.(\ref{eqn:UD_Lewis}) with eq.(\ref{eqn:UD_Ludwing}), 
it is found that these equations are identical 
when we set  
\begin{eqnarray}
    Q = R, \;\;\; P = R-K. 
    \label{transformation_parameters}
\end{eqnarray}
Therefore, Lewis model, eq.(\ref{eqn:Lewis}), brings about  
the same piecewise linear cusp bifurcation 
as Ludwig model, eq.(\ref{eqn:Ludwig}).
Note that the original continuous Lewis model 
has different terms from the original continuous Ludwig model.
However, the ultradiscrete equations for the Ludwig model and 
the Lewis model are essentially identical, 
and the common dynamical structure can be extracted and expressed 
with the same max-plus representation.
%

%

\section{Conclusion}
\label{sec:6}

In this paper, we have reported the investigation of the cusp bifurcation in the one-dimensional discrete dynamical systems derived via tropical discretization 
and in the max-plus dynamical systems obtained by ultradiscretization.
Based on our proposed proposition, we show that 
these dynamical systems retain the cusp bifurcation 
of the original continuous differential equation, 
using the Ludwig model and Lewis model as representative examples.
The cusp bifurcation point, curve, and surface of the tropically discretized systems coincide with those of the original differential systems. 
Furthermore, the ultradiscrete max-plus equations can also express 
the cusp bifurcation properties in a piecewise linear form.
We expect that the tropical and ultradiscrete approaches provide a promising framework for describing nonlinear and nonequilibrium phenomena with discrete and max-plus dynamical systems.
%


\bigskip

\noindent
{\bf Acknowledgement}

The authors are grateful to Prof. M. Murata, Prof. K. Matsuya, Prof. D. Takahashi, Prof. R. Willox, Prof. H.
Ujino, Prof. Y. Sato, Prof. A. Shudo, Prof. Emeritus Y. Aizawa, Prof. T. Yamamoto, and Prof. Emeritus A. Kitada for useful comments and encouragements.
This work was supported by JSPS
KAKENHI Grant Numbers 22K03442, 22K13963, 25K00212, and 25K07140.

\end{document}